\documentclass[sigconf, authorversion=true]{acmart}
\usepackage{multirow}
\usepackage{float}
\usepackage{stfloats}

\AtBeginDocument{%
  }

\copyrightyear{2025}
\acmYear{2025}
\setcopyright{cc}
\setcctype{by}
\acmConference[CHI '25]{CHI Conference on Human Factors in Computing Systems}{April 26-May 1, 2025}{Yokohama, Japan}
\acmBooktitle{CHI Conference on Human Factors in Computing Systems (CHI '25), April 26-May 1, 2025, Yokohama, Japan}\acmDOI{10.1145/3706598.3713646}
\acmISBN{979-8-4007-1394-1/25/04}

\makeatletter
\@ifundefined{textls}{\long\def\textls{\@ifnextchar[{\@textls}{\@textls[]}}
\long\def\@textls[#1]#2{#2}}{}
\makeatother
\begin{document}

\title[Exploring K-12 Physical Education Teachers’ Perspectives on Opportunities and Challenges of AI Integration]{Exploring K-12 Physical Education Teachers’ Perspectives on Opportunities and Challenges of AI Integration Through Ideation Workshops}

\author{Dakyeom Ahn}
\email{adklys@snu.ac.kr}
\orcid{0009-0004-9139-1605}
\affiliation{%
  \institution{Department of Physical Education}
  \institution{Seoul National University}
  \city{Seoul}
  \country{South Korea}
}

\author{Hajin Lim}
\email{hajin@snu.ac.kr}
\orcid{0000-0002-4746-2144}
\authornote{Corresponding Author}
\affiliation{%
  \institution{Department of Communication}
  \institution{Seoul National University}
  \city{Seoul}
  \country{South Korea}
}

\renewcommand{\shortauthors}{Dakyeom Ahn, Hajin Lim}


\begin{abstract}
While AI's potential in education and professional sports is widely recognized, its application in K-12 physical education (PE) remains underexplored with significant opportunities for innovation. This study aims to address this gap by engaging 17 in-service secondary school PE teachers in group ideation workshops to explore potential AI applications and challenges in PE classes. Participants envisioned AI playing multidimensional roles, such as an operational assistant, personal trainer, group coach, and evaluator, as solutions to address unique instructional and operational challenges in K-12 PE classes. These roles reflected participants’ perspectives on how AI could enhance class management, deliver personalized feedback, promote balanced team activities, and streamline performance assessments. Participants also highlighted critical considerations for AI integration, including the need to ensure robust student data security and privacy measures, minimize the risk of over-reliance on AI for instructional decisions, and accommodate the varying levels of technological proficiency among PE teachers. Our findings provide valuable insights and practical guidance for AI developers, educators, and policymakers, offering a foundation for the effective integration of AI into K-12 PE curricula to enhance teaching practices and student outcomes.
\end{abstract}

\maketitle

\section{Introduction}

Physical education (PE) plays a critical role in fostering students' holistic development by enhancing physical health ~\cite{strotmeyer2021effects} and overall well-being ~\cite{de2018effects}. Recognizing this importance, PE has become an integral part of compulsory education curricula worldwide ~\cite{hardman2014unesco}. However, consistently delivering high-quality PE remains a challenge due to limited resources, variations in curriculum implementation, and a heavy reliance on individual teachers' expertise ~\cite{landi2021physical, Powell2014Teachers}. The unique demands of PE—such as the need for extensive space, limited classroom hours, and the ability to accommodate diverse skill levels while ensuring safety—further complicate resource allocation and pedagogical innovation ~\cite{cothran2014classroom, sieberer2016effective}. Current approaches often fall short in addressing these multifaceted issues ~\cite{Rimmer1989Confrontation}, highlighting the need for novel and scalable solutions.

In this context, artificial intelligence (AI) has emerged as a potential resource for tackling these challenges by supporting teachers and facilitating personalized learning environments. In a wide range of K-12 education areas, AI integration has shown significant promise in enhancing teaching and learning outcomes ~\cite{roll2016evolution, yaugci2019prediction, o1986learning, holmes2023artificial}. For example, by leveraging AI-driven tools, educators have improved not only students' academic performance ~\cite{wu2013expert} but also their motivation and engagement with learning ~\cite{gavrilovic2018algorithm, azcona2019detecting, verner2020exploring, balakrishnan2018motivating}. Moreover, AI technologies have offered opportunities to monitor individual student progress ~\cite{wang2016course, pereira2020using, sapounidis2018latent} and provide personalized feedback ~\cite{lee2021applying}, enhancing the overall quality of education.

Despite these advancements, AI applications in K-12 education have predominantly focused on STEM subjects taught in classroom settings ~\cite{xu2022application}. However, the widespread adoption of AI in professional sports provides compelling evidence of its potential to support PE as well. For instance, AI tools have been employed to monitor and evaluate athletes’ skills ~\cite{bartlett2006artificial, ratiu2010artificial, fister2015computational}, optimize tactical decision-making ~\cite{lapham1995}, and analyze performance metrics to foster improvement ~\cite{walton2019exploring}. Commercial tools, such as Dartfish and BEPRO, illustrate how AI can deliver actionable insights for enhancing physical performance ~\cite{walton2019exploring}. These developments suggest that AI may hold promise for improving PE by addressing teachers' challenges and enhancing student engagement in K-12 PE classes.

Nevertheless, integrating AI into the context of K-12 PE presents distinct challenges. Unlike traditional classroom subjects, PE emphasizes physical activities conducted outside conventional learning environments ~\cite{zawacki2019systematic} and is often marginalized in school curricula ~\cite{corbin2021, harris2018case}. Furthermore, although AI’s potential in education has been widely acknowledged, its integration into practice is hindered by significant gaps, such as the absence of pedagogical frameworks tailored to PE ~\cite{Chen2020application, hinojo2019artificial, zawacki2019systematic, tang2023trends}. 

Addressing these gaps requires more than technological innovation; it demands a deeper understanding of the individuals who would be responsible for implementing these innovations in practice. As key facilitators of such educational change, PE teachers could play a critical role in shaping how AI can be practically integrated to enhance the quality of K-12 PE. Understanding their attitudes, expectations, and concerns toward AI would be essential ~\cite{nicholson2022participatory}, as it would bridge the gap between technological potential and the realities of the classroom. Such insights ensure that AI solutions align with pedagogical goals while addressing the unique challenges of PE ~\cite{johnson2022failure}. Furthermore, teachers’ perspectives could provide invaluable guidance on defining learning objectives and setting realistic expectations for AI integration in education ~\cite{Ertmer2012Teacher, Mama2013Developing}.

Recognizing the pivotal role of PE teachers in shaping the future of PE, this paper aims to explore the existing challenges PE teachers encounter in their classes and investigate how AI could be envisioned as a tool to address these challenges. Furthermore, we examine the key considerations necessary for effectively integrating AI into K-12 PE, ensuring alignment with both pedagogical goals and practical classroom needs. To guide our exploration, we posed the following research questions(RQs):

\begin{itemize} 
\item RQ 1. What challenges do PE teachers encounter in their classes?
\item RQ 2. How do they envision AI supporting them in addressing these challenges?
\item RQ 3. What concerns do PE teachers have regarding the adoption and implementation of AI in PE? 
\end{itemize}

To address these RQs, we conducted a focus group ideation workshop with 17 secondary PE teachers in South Korea. Participants highlighted key challenges, such as managing large classes and providing personalized feedback, and envisioned AI as a practical solution tailored to these demands. Specifically, they identified four potential roles for AI in PE: ``operational assistant'', ``personal trainer,'' ``group coach,'' and ``evaluator.'' These roles reflect AI's potential to assist with class management, deliver personalized feedback, facilitate balanced team activities, and streamline performance assessments. However, participants also expressed concerns about the integration of AI, including issues related to institutional support, varying technology proficiency among PE teachers, risk of AI overuse, and student data security. 

The contribution of this study is the following: 
\begin{itemize}
    \item Expanding understanding of AI's potential in the underexplored context of K-12 PE.
    \item Identifying multidimensional potential roles of AI that address both instructional and operational needs in PE classes.
    \item Highlighting ethical and practical considerations critical for successful AI implementation in education.
\end{itemize}

\section{Related Work}
In this section, we addressed several topics closely related to our research objective, including (1) the role and objectives of PE in K-12 settings, (2) recent developments in AI applications for student learning and teacher support, and (3) the importance of teacher involvement in designing AI-integrated curricula.

\subsection{Physical Education in K-12} 
According to the Society of Health and Physical Educators (SHAPE) in the United States, PE is an ``academic discipline designed to provide students with a planned, sequential, standards-based K-12 program of curricula and instruction to develop motor skills, knowledge, and behaviors for active living, physical fitness, sportsmanship, self-efficacy, and emotional intelligence'' ~\cite{america2015essential}. According to UNESCO's worldwide survey of school PE ~\cite{hardman2014unesco}, PE is typically a required or standard component of the K-12 curriculum in most countries during the compulsory education phase, reflecting the importance of PE in promoting physical activity and health among youth. Despite some regional and national variations, the general global trend indicates that both elementary and secondary schools allocate approximately 100 minutes per week~for~PE~classes.

The core curriculum models of PE encompass three main areas: movement education, which emphasizes basic motor skills as prerequisites for lifelong physical activity; sports education, which helps students become proficient in diverse sports; and fitness education, which promotes personal fitness and health maintenance ~\cite{2013Etsb}. In practice, many PE classes primarily endeavor to immerse students in a variety of physical exercises and sports ~\cite{chen2018toward, castelli2015contextualizing, Bar1995Health}. However, these classes also serve as platforms for honing social skills, thus bolstering social competence ~\cite{pangrazi2003physical, dishman2015motivation}.
Ultimately, PE aims to promote the holistic development of students, encompassing physical, mental, emotional, and cognitive abilities ~\cite{2013Etsb}. This comprehensive approach to student development is globally shared as a core objective in K-12 PE ~\cite{lynch2019physical}, reflecting the multifaceted nature of PE and its importance in overall student growth.

Given the multifaceted objectives of K-12 PE, the scope of content that PE teachers are expected to address in their classes is broad. Usually, nationwide standard curricula provide a guiding framework, but they are not binding. As a result, the overall curriculum structure is often left entirely to the discretion of PE teachers ~\cite{landi2021physical}. Additionally, the nature of PE, rooted in physical activity across large open spaces such as pools, soccer fields, gymnasiums, and outdoor courts, demands both specialized equipment and extended instructional time ~\cite{cothran2014classroom, Bevans2010Physical}. Furthermore, given the heterogeneous skill sets and abilities of many students, stringent control measures are crucial to ensure students' safety ~\cite{chepyator2003pre, o1994rules, siedentop2022introduction}. Consequently, achieving optimal PE outcomes is inherently challenging and depends significantly on PE teachers' classroom management and pedagogical strategies ~\cite{Rimmer1989Confrontation}. Specifically, students' PE learning outcomes largely hinge on their teachers' ability to effectively utilize space and resources and select instructional strategies that fit the classroom context and student characteristics ~\cite{Powell2014Teachers, sieberer2016effective}. Therefore, to address these challenges and support PE teachers in delivering high-quality education, it would be essential to actively involve them in the design and development of technologies, particularly when integrating advanced systems such as AI.

\subsection{AI for Education and Sports} 
Artificial Intelligence in Education (AIED) offers a wide range of opportunities for improving education quality and enhancing learning outcomes ~\cite{crompton2021potential, crompton2020psychological, chen2020artificial}. AIED applications broadly focus on supporting both students and teachers by providing tailored educational experiences and addressing various learning needs.

A large body of research has focused on enhancing student learning. For example, intelligent tutoring systems have been extensively researched, with the goal of providing personalized learning paths based on real-time analysis of learner behavior and performance ~\cite{vanlehn2011relative, ma2014intelligent}. Other notable applications include real-time analysis of student progress ~\cite{mirzaeian2016learning, roschelle2020ai}, and personalized feedback system on student work ~\cite{foster2019barriers, zhu2020effect}.

Simultaneously, AIED provides comprehensive support for instructional tasks, enhancing teachers' capabilities and efficiency. Research on automated summative assessment, for instance, focuses on improving the accuracy and reliability of AI-based automatic scoring systems ~\cite{hsu2021attitudes}. These research areas extend into classroom monitoring and orchestration, investigating how AI can effectively support teacher decision-making in complex classroom environments ~\cite{song2021review, vanlehn2021can, lee2024investigating}.

While AI applications have been extensively integrated into lecture-based classes in STEM disciplines ~\cite{bates2020can, kabudi2021ai}, there have also been some initiatives to employ data mining for sports training and skill analysis ~\cite{pan2019big}, as well as deep learning approaches to assess the quality of training ~\cite{wang2022evaluation} in PE contexts. For example, recent studies have demonstrated the effective use of AI technologies in training dance skills ~\cite{wang2024artificial, feng2022automatic} and rehabilitation ~\cite{jacob2021ai, vourganas2020individualised}. Building on these advancements, AI technologies integrated with tangible user interfaces~\cite{bakker2012embodied}—offer intuitive methods to enhance sports training by combining tactile and visual senses ~\cite{gallud2022use, nacher2015game}. 

Despite these promising applications, the scope of AI integration into general K-12 PE classes remains limited ~\cite{edtech_ai_k12_2023}. Current implementations often emphasize the development of specific sports skills rather than comprehensively addressing the multifaceted educational goals and the broader context of the PE classroom. To bridge this gap, this paper aims to explore opportunities for integrating AI into PE classes by identifying unique needs and challenges through focus group ideation workshops involving PE teachers.

\subsection{Challenges in Integrating AI into K-12: Importance of Teacher Involvement}
AI is driving educational advancements ~\cite{alberola2016artificial, chakroun2019artificial, park2016teachers} by promoting tailored learning experiences for individual students ~\cite{bernacki2020towards, samani2017bridging} and providing diverse ways to amplify the capabilities of educators ~\cite{ZHANG2021100025, edwards2018not}. Nevertheless, the integration of AI in educational settings presents complex challenges and considerations, given the multitude of stakeholders involved in the educational ecosystem. In particular, issues surrounding the collection, use, and protection of student data have persisted, especially concerning cybersecurity threats  ~\cite{grayson1978education, huang2023ethics}. Thus, students keep demanding transparency and control over their data usage, extending beyond mere information provision to actual data management rights ~\cite{jones2019learning, slade2014student, sun2019s}.

Also, recent AIED research consistently underscores the absence of a pedagogical context ~\cite{Chen2020application, hinojo2019artificial, zawacki2019systematic, tang2023trends}. AIED innovations often remain in the experimental or research phase. As a result, a significant gap exists between the potential capabilities of AIED and its actual implementation in educational settings ~\cite{bates2020can, kabudi2021ai}. Moreover, the successful adoption of AIED technologies depends on their alignment with learning objectives, educators' instructional strategies, and expectations regarding the role of such technologies ~\cite{Ertmer2012Teacher, Mama2013Developing}. Therefore, in order to effectively implement AIED systems, it is imperative to co-construct a ``clear and specific framework for innovation'' ~\cite{roschelle2006co} through the active involvement of key stakeholders such as teachers and reflect their educational practices and beliefs in the design processes ~\cite{Ertmer2012Teacher, Mama2013Developing}.

Therefore, an increasing number of AIED projects involved various stakeholders, such as teachers and students, in the design and development processes of educational technologies. This approach focuses on integrating the tacit knowledge from participants' real-life experiences with the knowledge of the researchers ~\cite{bodker2022participatory, mckercher2022beyond, schuler1993participatory}. Particularly, a recent study highlighted the importance of teachers' perspectives in incorporating AI training and tools into K-12 curriculum ~\cite{Lin2021Engaging}. For example, collaboration with teachers on the development of science education curriculum resulted in the successful implementation of the program ~\cite{durall2019co}. In light of this, this study engaged PE teachers in focus group ideation workshops to envision AI-integrated PE classes by incorporating their needs, challenges, and lived experiences.

\begin{table*}
  \centering
  \caption{Participant Demographic and Background Information}
  \label{tab:participant_info}
  \begin{tabular}{p{1.5cm} p{1cm} p{1.2cm} p{1.7cm} p{4.2cm} p{3cm}}
    \toprule
    \textbf{Group} & \textbf{ID} & \textbf{Gender} & \textbf{Teaching \newline Experience} & 
    \textbf{General Attitude Toward AI \newline(1: negative - 5: positive)} & 
    \textbf{Experience \newline Teaching with AI} \\
    \midrule
    \multirow{2}{*}{Group 1} 
      & P1  & M & 25 & 3.17 & No \\
      & P2  & M & 7  & 3.75 & No \\
    \midrule
    \multirow{2}{*}{Group 2} 
      & P3  & F & 18 & 3.42 & No \\
      & P4  & M & 10 & 4.33 & No \\
    \midrule
    \multirow{3}{*}{Group 3} 
      & P5  & M & 24 & 3.83 & Yes \\
      & P6  & M & 7  & 3.08 & No \\
      & P7  & F & 1  & 3.83 & No \\
    \midrule
    \multirow{3}{*}{Group 4} 
      & P8  & M & 9  & 3.25 & No \\
      & P9  & M & 10 & 3.83 & No \\
      & P10 & M & 11 & 3.75 & No \\
    \midrule
    \multirow{2}{*}{Group 5}
      & P11 & F & 8  & 4.08 & No \\
      & P12 & M & 9  & 3.58 & Yes \\
    \midrule
    \multirow{3}{*}{Group 6} 
      & P13 & M & 5  & 4.33 & Yes \\
      & P14 & M & 9  & 3.25 & No \\
      & P15 & F & 1  & 3.67 & No \\
    \midrule
    \multirow{2}{*}{Group 7} 
      & P16 & M & 5  & 3.25 & No \\
      & P17 & F & 2  & 3.75 & No \\
    \bottomrule
  \end{tabular}
  \Description{This table displays the participant demographic and background information of 17 participants. The data present the participants’ gender, teaching experience, general attitudes toward AI, and experience teaching with AI.}
\end{table*}

\section{Method} 
To explore the potential and challenges of integrating AI into K-12 PE classes, we conducted focus-group ideation workshops with 17 secondary PE teachers. Our approach drew inspiration from participatory design methodology ~\cite{spinuzzi2005methodology}, which positions users as experts with valuable knowledge of their experiences and context ~\cite{schuler1993participatory}. Through the ideation process, we aimed to elicit in-depth insights into the PE teachers’ needs, challenges, and their envisioned roles for AI in addressing the unique demands of PE. This approach allowed us to engage participants in open-ended, collaborative discussions, encouraging them to reflect on their experiences and creatively articulate their expectations for AI in PE.

Our focus group ideation session was divided into three parts. First, the in-service PE teachers, our participants, shared the specific challenges they faced in their classrooms and their perspectives on AI's general potential. Next, participants engaged in ideation and sketch activities, where they envisioned AI-integrated solutions for improving PE classes. Finally, the teachers identified the prerequisites, potential obstacles, and broader considerations necessary for integrating AI into PE. 

\subsection{Research Team Background}
Our study team conducted the study by combining diverse expertise and experience in HCI research and PE. The first author of our research team had received training as a PE educator and has an in-depth understanding and empathy for challenges associated with K-12 PE. This shared background fostered a foundation of trust and rapport with PE teachers during our focus group ideation workshop. Another author, with extensive experience in running design workshops with various stakeholders, played pivotal roles in shaping the design of the sessions, ensuring they were engaging and effective.

\subsection{Participants and recruitment}
To recruit in-service PE teachers, we posted an announcement in an online community of PE teachers and sent direct messages to some active members. In addition, we also contacted the alumni network from the Department of PE of the university with which researchers were affiliated and distributed the recruitment posting via their mailing list. The participants in the study were required to be in-service secondary school PE teachers who had passed a teacher certification examination. PE teachers who expressed willingness to participate in the study were encouraged to invite their fellow PE teachers. Through this process, seventeen secondary school PE teachers (12 males, 5 females) participated in this study, with an average teaching experience of 9 years (range: 1-25 years) (see Table 1).  All participants were Asian of South Korean nationality, and all study procedures were conducted in Korean.

\subsection{Procedures}
When participants reached out to us, we explained the detailed focus group ideation workshop and provided a link to a pre-survey online using Jotform. In the pre-survey, We asked participants to indicate their availability for scheduling and to provide basic demographic information, and general attitudes towards AI using the General Attitudes towards Artificial Intelligence Scale (GAAIS)~\cite{schepman2020initial}. 
This scale was implemented to gain an overall understanding of participants’ perspectives on AI prior to the workshop. We also gathered responses about participants' experiences with AI in PE, asking whether they had used any AI tools for their classes. A few days before the workshop date, we sent participants instructions on the focus group ideation session reminder (e.g., time, location, and expected duration) via text message. 

\begin{figure*}[!ht]
  \centering
  \includegraphics[width=506pt]{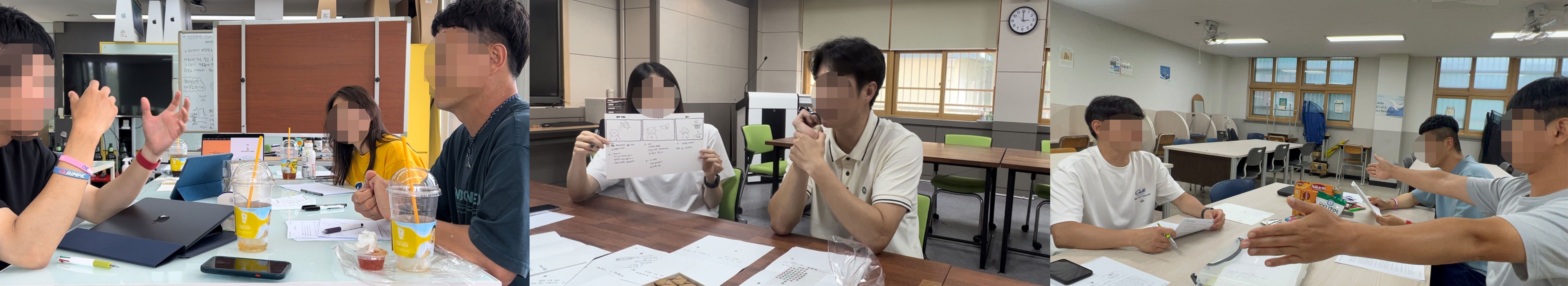}
  \captionsetup{justification=centerlast, singlelinecheck=false}
  \caption{Focus group ideation workshops with secondary school PE teachers. \\Left: ``Crazy 4's'' idea generation session. Middle: storyboard development activity. Right: Role-playing exercises.}
  \Description{The three photos show images of secondary school physical education teachers engaged in the process of problem-finding and solution-sketching. The left features 'Crazy 4's' brainstorming. The center shows two discussing their storyboard results, and the right features a role-play.}
  \label{fig:fig1}
\end{figure*}

All sessions were conducted in person, with each session consisting of two or three PE teachers grouped together (see fig. \ref{fig:fig1}). This small group format was chosen for active interaction among participants, trust building, and collaborative group thinking ~\cite{slingerland2022participatory}. While we maintained a neutral stance to avoid biasing participants' responses, we actively managed speaking opportunities to ensure that all participants were able to contribute their perspectives.

Our focus group ideation workshops were divided into four activities: \textbf{1) introduction}, \textbf{2) focus group discussion} to identify challenges in running PE classes, \textbf{3) solution ideation and sketching session}, and \textbf{4) wrap-up focus group discussion}, which took about 2.5 hours. The study procedures were approved by the ethics institutional review board (IRB) of Seoul National University, where the study was hosted.

\subsubsection{Introduction: 10 minutes}
In each session, participants were fellow PE teachers at the same middle schools or high schools, except for Group 3. After all participants had entered the meeting room, the first author shared the purpose of the research and obtained signed consent. All sessions were audio recorded, and we digitally captured all the artifacts created during the sessions. Participants were then informed that they could stop the session or take a break at any time. Also, we encouraged participants not to feel pressured to answer unfamiliar questions, as participants were not experts in AI technologies. In this way, we tried to create a comfortable environment where participants could share their PE class experiences and actively participate in workshop activities. 

\subsubsection{Focus group discussion: 20 minutes}
In this phase, we aimed to uncover the challenges participants face in their PE classes and create a foundation for shared understanding to inform the subsequent ideation session. The discussion began with participants reflecting on specific challenges they encountered while managing and running PE classes. 
Throughout this phase, participants were encouraged to express their thoughts openly, share their experiences, and engage collaboratively to build on one another’s ideas, ensuring a collective understanding of the context and issues.

\subsubsection{Solution Ideation and Sketching Stage: 1.5 hours}
In this phase, participants were given a brief lecture on the primary capacities and foundational principles of AI, based on existing AI literacy education resources ~\cite{Kim2020ai, google-oii-atozofai}. The content was divided into four main categories of AI applications:  \textit{perception}, \textit{natural language processing} (NLP), \textit{automation and optimization}, and \textit{reasoning and prediction}. Each category included explanations of the concepts of tasks possible through AI and the main application cases currently being widely utilized. To reduce unfamiliarity and enhance understanding, a few key examples closely related to daily life were covered in depth. These examples focused on explaining what kind of AI each application represented, how it worked, and what problems these systems tried to solve. This approach was intended to provide participants with a basic yet comprehensive understanding of AI, enabling them to effectively brainstorm and formulate concrete ideas for incorporating AI into PE classes.

Specifically, the \textit{perception} section focused on how AI mimics human senses to gather and interpret data, exemplified by computer vision's ability to recognize objects or events in images and videos. The \textit{NLP} section explained how AI processes meaning and context from text and voice data, enabling functions like conversation, summarization, and translation. In the \textit{automation and optimization} part, the emphasis was on AI's role in automating repetitive tasks and aiding in complex decision-making, with autonomous cars as a main example. Finally, the \textit{reasoning and prediction} part centered on AI's capability to draw conclusions and make predictions based on data, illustrated using the prediction of baseball game outcomes. During the lecture, the researcher monitored participants' understanding of the material and actively encouraged them to ask questions if they were unsure about anything.

Notably, our lecture content did not directly address the potential harms and biases of AI, as we believed that introducing these aspects upfront could limit the scope of their ideas. Instead, discussions about potential AI harms emerged naturally as participants shared and explored their ideas throughout the workshop. This approach allowed for a more organic exploration of AI's challenges and limitations, grounded in the participants' own understanding of AI and their specific PE class contexts.

After each lecture, participants were asked to individually sketch four ideas for applying this specific AI capability to their PE classes. We modified the \textit{Crazy 8's} ideation technique ~\cite{GoogleDesignSprints2023} into a \textit{Crazy 4's} approach, where participants focused on generating four ideas instead of eight.
Based on this, participants were instructed to sketch one idea per minute on a quartered A4 sheet of paper. We explained that their sketches could encompass anything from textual descriptions of the system's primary functions to conceptual illustrations. During the brainstorming process, we consciously avoided placing constraints on the feasibility of their solution ideas. It was because our primary focus was to gather insights centered on the needs and contexts of our research subjects rather than focusing on the technical feasibility of implementing these solution ideas. 

After all participants completed sketching, each participant shared their ideas. Among the ideas presented, each participant was asked to vote for two ideas they believed would be most beneficial in PE classes and explain their reasons for choosing those specific ideas.

The process of `Short lecture - Sketching four ideas - Voting' was repeated four times, once for each of the four AI categories. This led to each participant creating a total of sixteen idea sketches and each group choosing eight top-voted ideas. From these eight top-voted ideas, participants engaged in discussions to select the final two ideas that they would like to develop further.

For each of these final top two ideas, participants were instructed to develop the ideas and complete a storyboard through group discussions. The primary goal of this storyboard creation activity was to encourage participants to think critically about the benefits and potential challenges of incorporating this specific idea into their actual classes. To facilitate this process, we provided them with the storyboard template, including boxes for drawings and blank spaces for descriptions, which consisted of three parts: \textit{problem situation,} \textit{solution}, and \textit{expected outcome}. Building on the storyboard, brief role-playing exercises were conducted to gain a more nuanced perspective on specific ideas. Each participant took on at least one role depicted in the storyboard (e.g., PE teacher, students, AI, administration staff, etc.) and acted out scenes from the storyboard.

\subsubsection{Wrap-up focus-group discussion: 20 minutes}
Finally, each group was prompted to discuss the requirements for successfully integrating AI into PE classes, drawing on previous activities. In particular, we asked participants to discuss potential challenges and considerations in integrating AI into PE classes. Specifically, we encouraged participants to focus on the realistic difficulties and concerns they anticipated in the development, adoption, and operation of AI-integrated PE classes. After the session, we compensated each participant with 80,000 Korean Won (approximately 60 USD).

\subsection{Data Analysis}
\begin{figure*}[htbp]
  \centering
  \includegraphics[width=460pt]{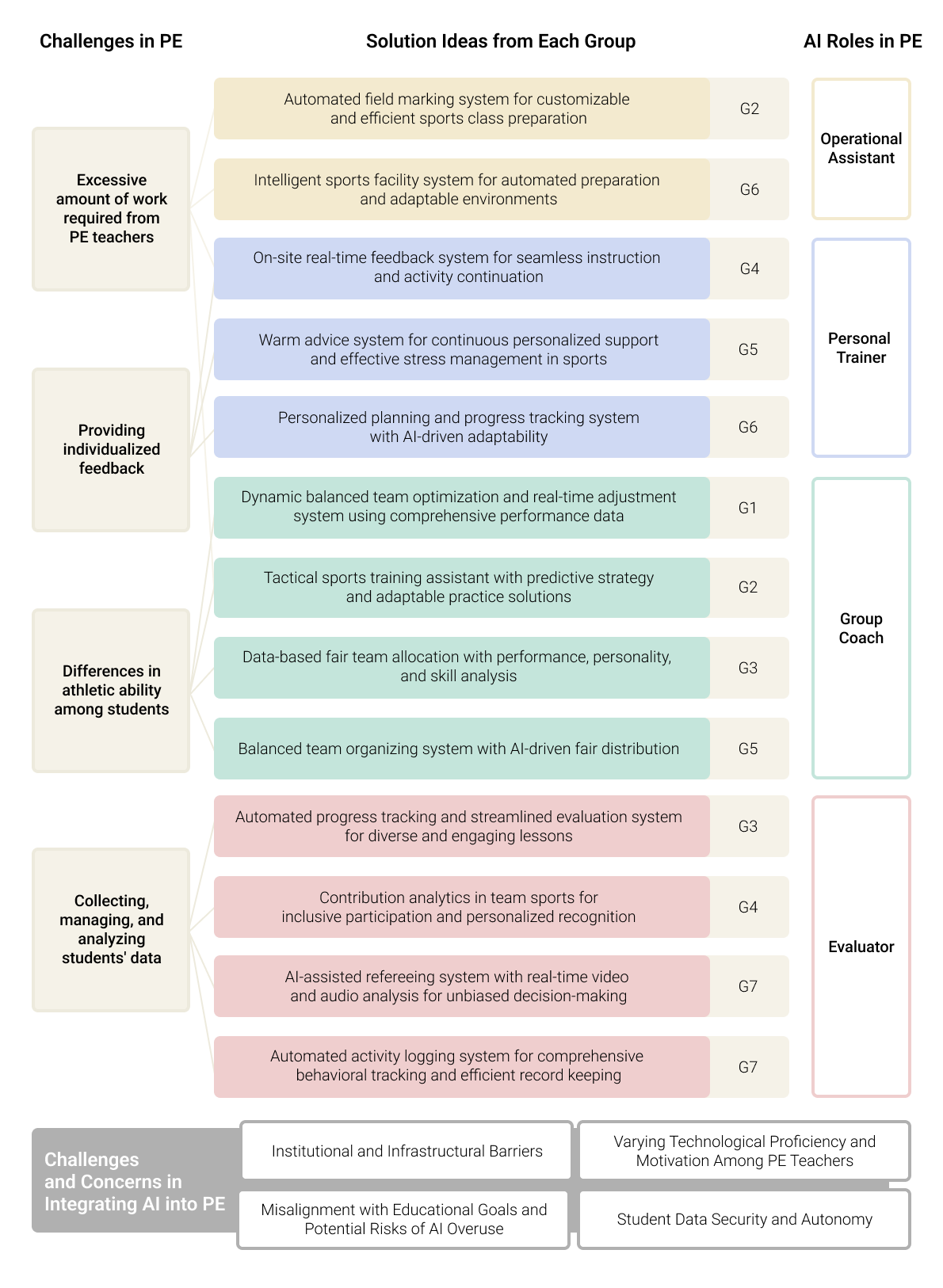}
  \caption{Thematic analysis map of findings}
  \Description{Thematic analysis map illustrates challenges in classes, AI roles, and considerations for AI integration in physical education. It outlines key challenges such as teachers' workload, the need for personalized feedback, disparities in athletic ability, and student data management. Proposed AI roles include Operational Assistant, Personal Trainer, Group Coach, and Evaluator. The map also addresses concerns regarding infrastructure limitations, teachers’ technological proficiency, alignment with educational objectives, and student data privacy.}
  \label{fig:fig2}
\end{figure*}

All sessions were video and audio recorded, and the first author transcribed the audio recordings. All participants' idea sketches and storyboards were digitally captured after the session and integrated into the corresponding group's transcripts. Using thematic analysis~\cite{patton2014qualitative}, two researchers individually read the transcripts multiple times and generated codes, resulting in a total of 1852 codes. Next, by identifying frequently occurring types of codes, common themes were iteratively identified, allowing for the development of higher-level themes. This process involved using NVivo software ~\cite{nvivo}, which facilitated an organized and systematic approach to categorize initial codes to high-level themes based on their thematic relevance. Following multiple rounds of discussion, we grouped the themes and codes into four broader categories: ``Challenges in PE,'' ``Solution Ideas from Each Group'', ``AI Roles in PE'' and ``Challenges and Concerns in Integrating AI into PE'' (see Fig. \ref{fig:fig2}). Participants' quotes were translated from Korean to English by bilingual English-Korean speakers.

\section{Findings}
According to the pre-session survey results, participants' attitudes towards AI were generally neutral but slightly leaning toward the positive side, with an average rating of 3.66 out of 5 and a standard deviation of 0.37, indicating a relatively small variation in attitudes among participants. Reflecting on this, all participants displayed a generally open-minded attitude toward the topic of the workshop sessions. Notably, 14 out of 17 participants had no prior experience using AI in their classes, underscoring the fact that AI had not yet been actively integrated into PE classes at the time of the study. 

Overall, our focus group ideation workshops were well-received, with all participants actively engaging in the focus group ideation activities. Some participants shared that the session had been engaging and facilitated active sharing of ideas and opinions: ``\textit{I was pleasantly surprised by our creative and practical ideas, which made this session both useful and enjoyable}'' (P7).

Throughout the brief lectures on AI's four key areas of capacity, participants actively engaged to enhance their understanding of AI. For example, P13, upon hearing about perception capabilities in AI, commented, ``\textit{Then, based on collected data, AI can learn the patterns and identify errors},'' showing his grasp of AI's ability to extract and analyze complex datasets, recognize patterns, and detect errors. Similarly, P17 demonstrated an understanding of the NLP topic, relating it to well-known technologies like ChatGPT. P4 further exhibited an understanding of AI's capabilities, stating, ``\textit{AI could potentially assist in analyzing students' movements, which might help provide more targeted feedback in certain situations,}'' reflecting a practical understanding of AI's diverse applications in PE settings. Despite limited AI expertise, participants engaged productively in discussions, proposing specific use cases with an open attitude. 

Overall, all participants agreed to the potential for AI \mbox{to enhance} PE classes: ``\textit{Using AI in PE classes might make them more effective or interesting}'' (P4). However, they emphasized that AI should play a ``\textit{supporting role}'' in their PE classes: ``\textit{AI should be an addition, not the main thing}'' (P6). P14 added ``\textit{It is best for AI to take the supporting role.}''

In the following sections, we first explore the challenges PE teachers faced in their classes and how participants envisioned AI addressing these challenges through taking various assistive roles. We then delve into participants' concerns regarding the challenges and potential risks of integrating AI into PE.

\subsection{Common Challenges that PE Teachers Face}
Participants reported several notable challenges in managing PE classes, including excessive workload, difficulty in providing individualized feedback, and issues arising from variations in students' athletic abilities and interests. Additionally, they struggled with the complexities of collecting, managing, and analyzing student data in evaluations.

Some participants (n=7) mentioned \textbf{excessive amount of work required from PE teachers}, given the short class duration. They said it was challenging to monitor and control students during PE classes since they were held in large, open spaces, such as sports fields and gyms: ``\textit{PE class differs from other subjects because it takes place outdoors, so you must continuously monitor and control your students}'' (P1). 
As they had a primary responsibility to ensure a safe learning environment for their students, they also had to closely examine and maintain the equipment used during the class: ``\textit{Equipment management is another concern. No matter how good the equipment is, if it gets hit by a ball, for example, you can't use it in the next class}'' (P17). As a result, many participants felt overwhelmed due to the class logistic tasks beyond just running lectures during class time: ``\textit{Even between classes, I sometimes have to rush to prepare for the next class}'' (P9).  

Another important challenge participants encountered was \textbf{providing individualized feedback} to students through understanding their learning profiles. However, teachers highlighted the challenges in PE classes where one teacher must instruct multiple students at the same time: ``\textit{Teaching 29 students in one class, even if I give feedback in groups for 5 minutes each, it's difficult to provide feedback to everyone. It is almost inevitable that some students will not receive any feedback}'' (P15). 

Also, there were shared experiences of challenges caused by \textbf{differences in athletic ability among students}: ``\textit{Teaching students with varying levels of athletic ability is really challenging}'' (P3). Particularly, nearly a third (n=6) also pointed out that this sometimes involved general physical differences between genders: ``\textit{Sometimes it is more effective to have classes separated by gender. It is because it is hard to facilitate game activities in mixed-gender classes due to physical differences between male and female students}'' (P3). P13 further elaborated, ``\textit{In activities focused on ball sports, female students tend to be somewhat marginalized},'' while P1 reported the opposite case: ``\textit{In activities that involve making something, girls usually excel and participate more actively, whereas boys are more passive than girls}.'' As a result, they reported difficulty in forming balanced teams for team sports and group activities: ``\textit{Even when the teacher tries to create the balanced team, students often complain, saying `the balance is off,' `it's not fun,' or `this is not fair' }'' (P2). 

Lastly, they highlighted the challenges associated with \textbf{collecting, managing, and analyzing students' data}. Specifically, P14 mentioned the difficulties in collecting student data while running classes: ``\textit{When running PE classes outside, it's not easy to assess and keep records of each student's record}'' (P14). Further, P17 underscored the challenges in managing data: ``\textit{As most of the records are usually hand-written, it takes at least two hours just to input recorded results into the computer}'' (P17). Further, P5 mentioned that since it became necessary to evaluate not only the outcome but also the progress, keeping records and analyzing them became more complex and burdensome: ``\textit{Although the focus has shifted towards process-oriented from final performance-based evaluations, it's difficult to collect and check the data of multiple students every day. Even with those data, I have difficulty organizing them into meaningful evaluations}'' (P5). 

\begin{figure*}[!hb]
  \centering
  \includegraphics[width=460pt]{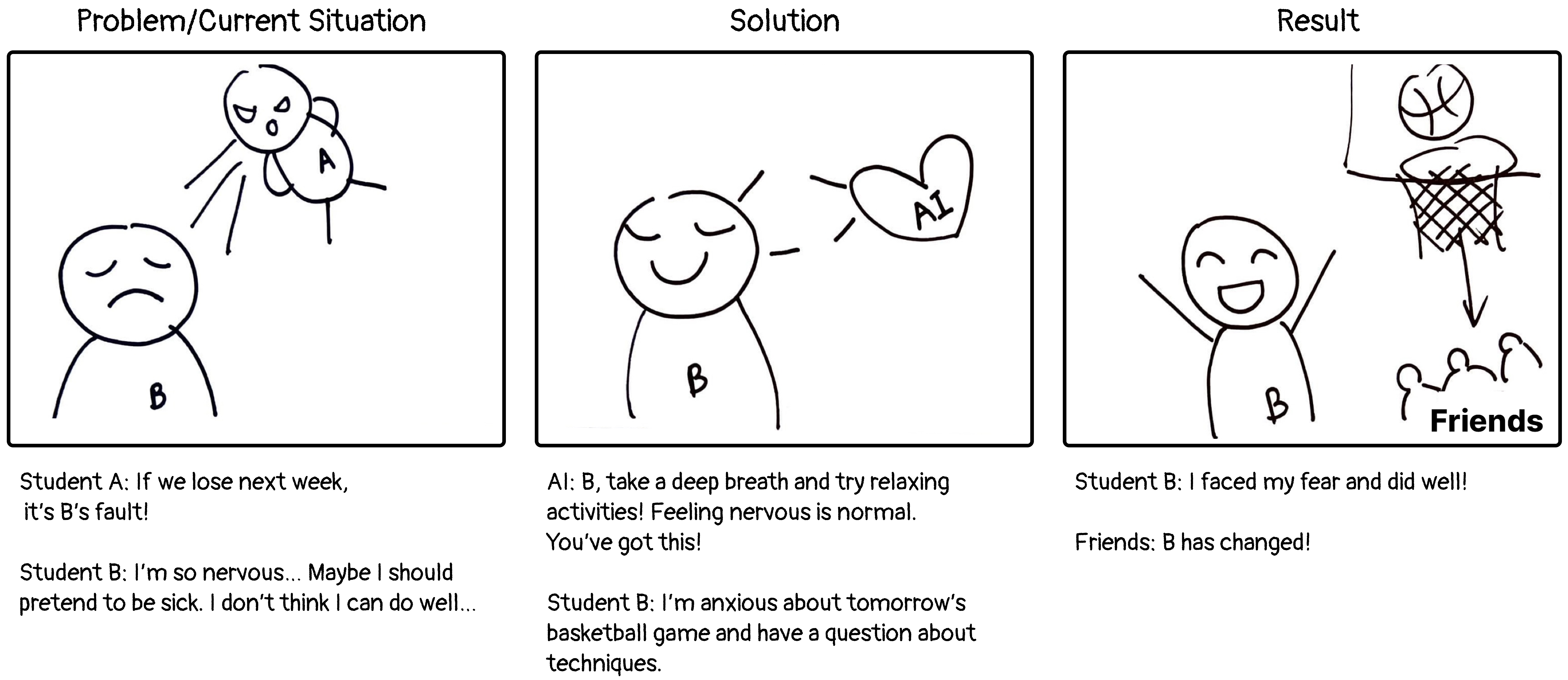}
  \caption{``Warm Advice AI'' idea storyboard from Group 5}
  \Description{The three boxes in the storyboard of group 5 are problem and current situation, solution, and result. The problem and current situation show the student getting nervous before a game and his friends blaming him. The solution shows the AI counseling the student to resolve their emotional issues. The result shows the students overcoming their nerves and having fun in the game.}\label{fig:fig3}
\end{figure*}

\subsection{Envisioning AI's roles in PE Class}
Based on the challenges they shared, participants generated a range of solution ideas where AI could play a role in addressing these issues. From their ideation outcomes, four overarching categories of expected roles of AI emerged: operational assistant, personal trainer, group coach, and evaluator, reflecting participants' collective vision of how AI could support diverse tasks and meet the unique demands of PE classes.

\subsubsection{AI as Operational Assistant}
All participants envisioned AI playing a supportive role in class operations, particularly by addressing excessive workloads and the challenges of managing large, open spaces. By assisting with these tasks, AI was seen as a tool to enable teachers to focus more on meaningful guidance and personalized feedback for students. Particularly, participants identified AI’s contributions to class operations in three key areas: environment setup, student management, and safety management. 

In the first domain, 11 participants proposed AI solutions that would optimize the class environment, such as lighting, temperature, and music: ``\textit{The gymnasium is often very noisy. Therefore, we need systems that can adjust noise levels, temperature, humidity, and sound level of music}'' (P13). P6 also had a similar idea: ``\textit{If AI could sense the overall temperature and adjust the air conditioning based on how much the students move, that would be helpful.}'' 

Moreover, three participants wanted AI to assist in student management tasks. They expressed a desire to leverage AI's computer vision technology to check students attending the class, not only to monitor attendance but also to check basic health status: ``\textit{It would be nice to have a system that can count the students to ensure everyone is present and check if anyone has a fever}'' (P3).

Also, seven participants proposed ideas for safety management. They wanted AI to detect and warn of any potential hazards in the class environment: ``\textit{I hope for preventive measures such as warning sounds if students engage in dangerous behaviors in certain places or situations}'' (P10). P7 also had a similar idea: ``\textit{In a flying disk class, there are many instances where students run after the disk and collide; it would be great if AI could detect this risky behavior in advance and sound an alarm.}''

Participants believed that getting operational help through AI solutions would allow PE teachers to have more time for meaningful guidance during classes: ``\textit{I think it would greatly reduce time constraints as AI can partially substitute for the teacher}'' (P10). P3 also mentioned the benefit of time-saving through AI: ``\textit{It's crucial to enable the PE teacher to invest the saved time into the students.}'' Also, they anticipated that the teaching efficiency would improve: ``\textit{I think we can teach students more things within the given time}'' (P10).

\subsubsection{AI as Personal Trainer}
The challenge of providing individualized feedback in large PE classes highlights AI's potential as a personal trainer. The majority of participants (N=13) envisioned it as a personal coach offering tailored guidance and support. Particularly, seven participants wanted AI to recognize students' physical activities, analyze the accuracy of their posture, and provide individualized feedback: ``\textit{Individual action analysis to give feedback on things like plank posture or arm angle would be good}'' (P14). 
Specifically, P13 hoped that AI could help with the timing of basic actions: ``\textit{When rotating shoulders or hips, it would be helpful if AI could analyze and signal the optimal timing to exert force or release.}'' Also, P2 imagined a system that could indicate optimal step locations for time-sensitive sports like the 100m dash: ``\textit{Considering different step locations for individual students, it would be great if the system could show where to step on for better performance}'' (P2). 

Also, almost half (n=8) of participants envisioned that AI could serve as a practice partner that assists students based on their physical fitness level. For example, P1 suggested an idea of an AI practice partner that could adjust the difficulty of techniques dynamically: ``\textit{AI would serve as a practice partner, giving slightly challenging missions depending on each student's level}'' (P1). 

Moreover, seven participants suggested ideas that analyzed real-time environmental data and provided tailored feedback to students: ``\textit{If it is windy during an outdoor flying disc lesson, it would be great if AI could give specific instructions like 'Throw a few meters to the right due to northeast winds at X speed'}'' (P12). 
In doing so, they expected AI to analyze a student's performance and environment data and provide appropriate guidance tailored to provide a personalized training experience.

While most ideas for AI as a personal trainer focused on the physical and technical aspects of PE, Group 5 envisioned a unique concept: the ``Warm Advice AI'' (see fig. \ref{fig:fig3}). This idea revolved around an AI system designed to provide psychological encouragement to each student, distinguishing it by addressing the emotional and motivational aspects of PE, in addition to the physical training: ``\textit{If AI provides warm, consistent, encouraging feedback to students who are nervous about competitions, it might help them overcome psychological pressure. Unfortunately, we can't do this for everyone during the class}'' (P11). Their idea reflected their expectation that AI could support both the emotional and physical needs of students in PE, offering a holistic approach to personalized support.

\begin{figure*}[!ht]
  \centering
  \includegraphics[width=460pt]{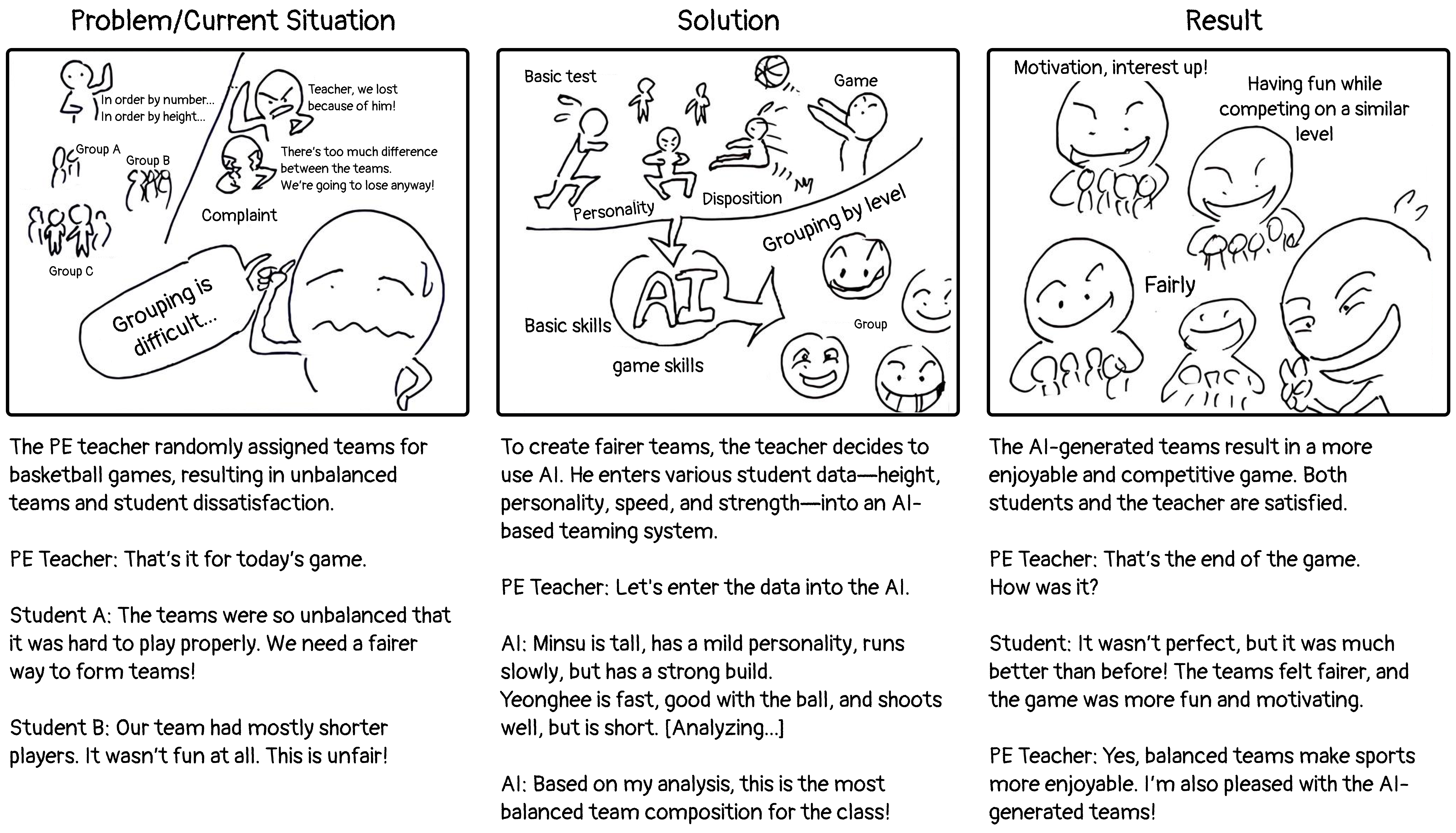}
  \captionsetup{justification=centering}
  \caption{``Automatic teaming AI'' idea storyboard and role-playing script from Group 3}
  \Description{Group 3's storyboard and role-playing script have three sections: 'Problem and Current Situation', 'Solution', and 'Outcome'. Initially, it shows students struggling with imbalanced teams and personality conflicts, resulting in weak team performance, despite the teacher's attempts. The solution involves using AI to analyze interactions and form balanced teams. The outcome demonstrates increased student engagement, easier team-building for the teacher, and improved student teamwork.}
  \label{fig:fig4}
\end{figure*}

\subsubsection{AI as Group Coach}  
Four of the seven groups developed storyboards in which AI served as a group coach, highlighting the challenge of balancing varying athletic abilities and interests among students. These storyboards envisioned AI enabling balanced team compositions and promoting smoother in-class team activity operations. Over half (n=9) of participants thought of AI  systems to create balanced team compositions based on students' data: ``\textit{By carefully analyzing available data on students' performance and attitudes, AI could assist in suggesting more balanced team compositions, which might lead to more engaging games}'' (P2). P7 also proposed a similar idea of analyzing the personalities of the students to form teams that could create the best synergy: ``\textit{By analyzing students' practice patterns, personalities, and cooperation attitudes,  AI could create teams of students with compatible personalities}'' (P7). 

Further, Group 4 imagined a tactical analysis system utilizing AI algorithms for complex game data analysis, recognizing offensive/defensive patterns, and aiding in strategic decision-making: ``\textit{AI might be able to assist in analyzing team data and suggest possible tactical approaches, which could complement the teacher's expertise in developing training strategies and game plans}'' (P4).

Using such AI solutions, they expected to create more balanced, welcoming, and inclusive classes. Specifically, they expected this could help create a class that both male and female students would be satisfied with: ``\textit{This would help the goal of my class to consider gender balance and create a class that increases satisfaction among different student groups.}'' (P2). Ultimately, they wished the AI solution could promote cooperative experiences in PE classes (see fig. \ref{fig:fig4}): ``\textit{PE classes frequently involve team activities, so fostering positive teamwork experiences is crucial.}'' (P6).

\subsubsection{AI as Evaluator}
\begin{figure}[hb]
  \centering
  \includegraphics[width=130pt]{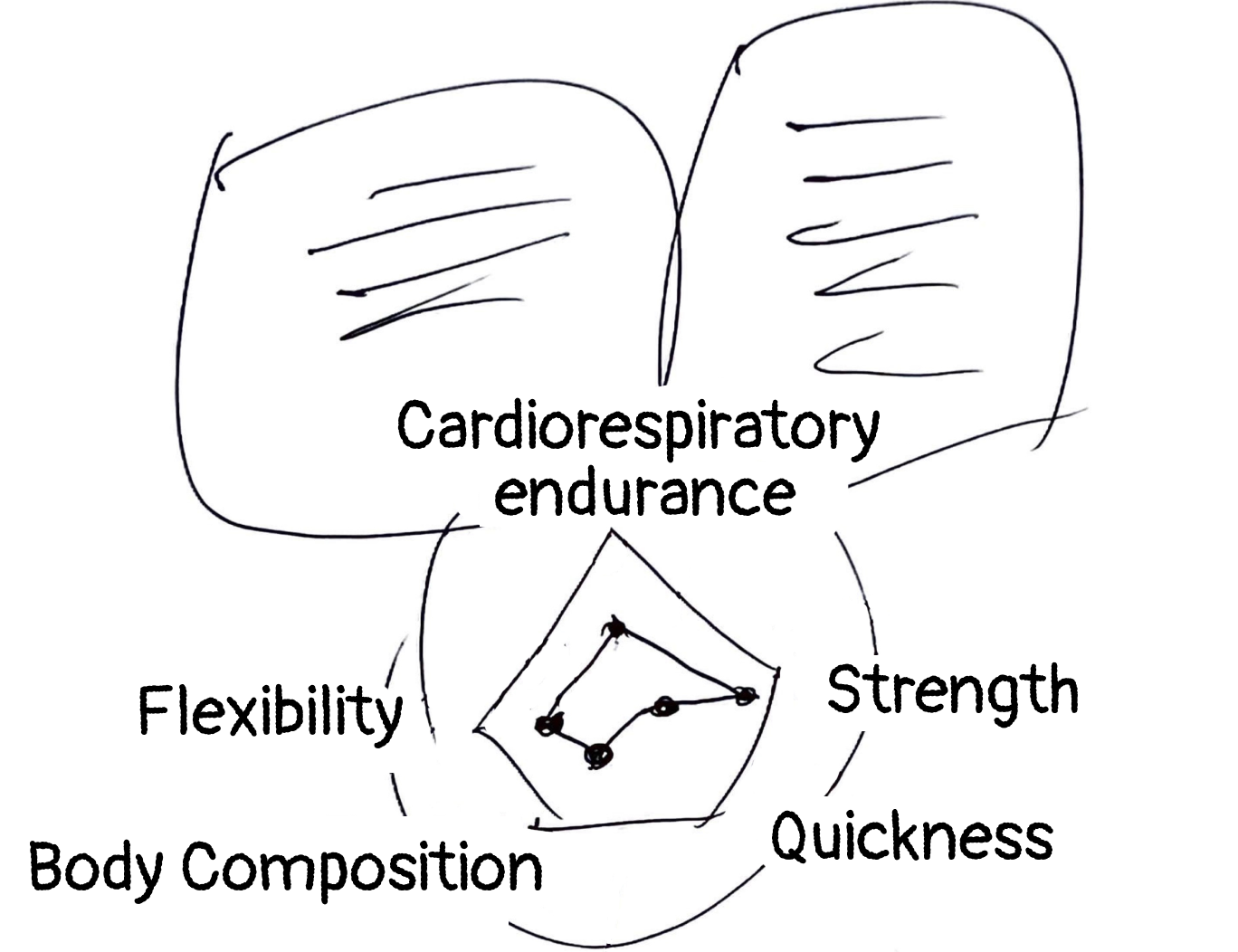}
  \captionsetup{justification=centering}
   \caption{P15's idea sketch for evaluating students in multidimensional ways using AI}
   \Description{The Crazy 4's results drawn by P15 illustrate the need to record and store student data. The image depicts an AI analyzing a student's baseline fitness.}
   \label{fig:fig5}
\end{figure}

Almost all (n=15) participants envisioned AI as an evaluator, addressing the challenges of collecting, managing, and analyzing students' data. They expected AI to streamline multi-dimensional evaluations by tracking both outcomes and progress in PE classes while analyzing various aspects of students' athletic performances and attitudes (see fig. \ref{fig:fig5}). 

Essentially, they expected AI to collect basic quantitative data like pass success rates and contributions to teamwork. For example, P6 imagined an AI system that automatically collected each student's in-game performance records: ``\textit{It would be helpful if basic data could be collected more efficiently during games, perhaps with some level of automation where feasible. For example, the number of times a student caught a ball, the number of steals, and so on.}'' They envisioned that such data could be collected via wearable sensors, equipment sensors, or motion capture technologies and then meaningfully processed by AI to track each student's performance.

Fundamentally, they wished for AI to streamline the data collection and management for multi-dimensional evaluation for each student: ``\textit{If data can be systematically collected and processed, we might be able to better understand and evaluate the outcome and the process, and potentially track the personal growth of each student.}'' In particular, many participants (n=7) envisioned AI solutions that could monitor students' language usage during a game as a way to monitor and assess students' classroom behavior and attitude: ``\textit{If a voice recognition works well, NLP algorithms could count the number of times students use swear words. In that way, we can see and discipline those who used such words many times}'' (P10). 

By streamlining activities involved in the progress-oriented and multi-dimensional evaluation processes, they expected to create a constructive class culture where ``\textit{Hard-working students are acknowledged as well, even when they are not in sight,}'' (P6) and ``\textit{both physical and mental aspects of students could grow}'' (P10).

\subsection{Challenges and Concerns in Integrating AI into PE}
Despite optimistic expectations regarding AI’s supportive roles in PE, participants raised various concerns about the challenges of implementation and the potential harms of AI integration. These apprehensions stemmed from the various challenges and obstacles they faced when adopting digital technology other than AI into their PE classes. 

\subsubsection{Institutional and Infrastructural Barriers}
The majority of participants (n=11) expressed concerns regarding \textbf{institutional support and resource allocation} for implementing AI in PE. These concerns were heightened by the fact that PE is considered a lower-priority subject in South Korea compared to other subjects, making the adoption of AI in PE potentially more complex. P5 pointed out that this obstacle was beyond teachers' capabilities and depended on broader institutional priorities and resources devoted to PE. He remarked, ``\textit{Even getting a basic whiteboard installed in the gym can be a hurdle. Given PE's diminished standing, [AI integration] might take forever.}'' In addition, P6 pointed out that increased technology use might lead to additional responsibilities: ``\textit{The more technology we introduce, the more work it creates for us. They need people to manage it, someone always has to be there to oversee it, and it inevitably generates extra tasks}'' (P6). 

Particularly, eleven participants identified \textbf{infrastructural and logistical issues} as major barriers to integrating AI in PE. These challenges included not only spatial constraints but also inadequate IT infrastructure (e.g., lack of Wi-Fi access) in typical PE settings such as school fields and gyms. P11 specifically addressed spatial concerns: ``\textit{For instance, if equipment like speakers or motion analyzers have to be installed, there is a lack of adequate space for these installations. Moreover, if the equipment is large, it reduces the space available for physical activity}.''

Additionally, four participants mentioned \textbf{technology maintenance}. P10 also mentioned: ``\textit{Current PE classes will face difficulty adopting AI technologies as they require devices such as tablets, which can easily get damaged.}'' Accordingly, they emphasized the importance of robustness and usability in AI systems. For example, P4 stated, ``\textit{If AI is not durable or easy to use, it won't deliver value worth the time and money invested in its development.}'' 

\subsubsection{Varying Technological Proficiency and Motivation Among PE Teachers}
The primary reflection revolved around the \textbf{limited technology proficiency among PE teachers} in employing AI. P11, reflecting on the COVID-19 pandemic era, noted that not all school teachers were naturally drawn to adopt digital tools unless external circumstances required them, noting, ``\textit{The reason the remote class was possible during the COVID-19 time was because the situation necessitated it, so we had to do that.}'' P9, who admitted not being tech-savvy, recalled challenges with remote teaching and spoke about the discomfort that he and other teachers experienced: ``\textit{In my school, there are a lot of senior teachers, and many feel overwhelmed by the rapid pace of technological shifts.}'' As such, almost half (n=8) of the participants voiced concerns about the disparity in tech proficiency among PE teachers: ``\textit{The pandemic underscored the tech divide, with many teachers feeling so outpaced that they considered early retirement. Especially in PE, a significant number of teachers remain rooted in traditional teaching methods and are hesitant about the introduction of new tools.} (P6).'' 

In a similar vein, another major consideration was the \textbf{limited motivation and initiative} among teachers to integrate AI into their classrooms. P13 highlighted, ``\textit{The teaching profession currently lacks incentives or rewards for educators who try to learn and implement AI technologies in their practices. At the same time, there are no disadvantages for not engaging with these new technologies.}'' Similarly, P9 pointed out, ``\textit{It's challenging for teachers to step forward because their efforts and achievements are not recognized, and they are not supported when issues arise from these attempts.}'' Additionally, P8 expressed concerns about the psychological burden associated with adopting new tools and AI systems, highlighting the additional learning and preparation required from teachers. He remarked, ``\textit{If we have to learn and research how to use specific programs constantly, it will undoubtedly create extra burdens.}'' 

\subsubsection{Misalignment with Educational Goals and Potential Risks of AI Overuse} 
Six participants were concerned that AI's introduction could \textbf{adversely affect PE classes' educational outcomes} if it did not align with the PE's educational goals: ``\textit{AI can provide convenience, but if we lose sight of the purpose of education, it is essentially meaningless}'' (P13). For example, P13, recalling an experience with a VR-integrated PE class, observed a reduction in actual physical activity time.  P5 also cautioned: ``\textit{AI needs to be used in alignment with PE's goals and philosophy. Using it solely for fun or curiosity may backfire.}'' Accordingly, they emphasized the importance of clear and cohesive instructional guidelines that could help them align the goals and practices of PE with the integration of AI:``\textit{A coherent curriculum and structured guidelines are necessary to harness AI effectively }'' (P13). 

Concerns were raised about the \textbf{potential overuse of AI} in PE group activities. P10 worried that excessive AI involvement might lead to overly structured classes, diminishing the inherent joy and dynamic aspects of sports: ``\textit{It's possible that if AI gets too involved, sports might lose their enjoyment if they are conducted strictly according to the past data}'' (P10). In addition, P10 expressed worry about students potentially over-relying on AI's instructions about AI's personalized feedback approach: ``\textit{If students start saying `But AI told me to do it this way' when I give feedback, or `The AI says to do it like this' when we're trying to move together as a group, it could lead to an overreliance on the AI}'' (P10). P13 also highlighted the potential adverse effects on students' self-directed learning: ``\textit{There's a concern that student autonomy might decrease. If everything is tailored and provided for them, they might not actively seek out what they need or try to figure things out on their own, potentially reducing their self-directedness}'' (P13). 

\subsubsection{Student Data Security and Autonomy}
Four participants expressed concerns about AI integration from students' perspectives. In particular, they stressed the need for \textbf{legal and ethical guidelines when collecting student data} for AI: ``\textit{To use a student's data for enhancing learning, explicit consent from both the student and parents will be necessary. Therefore, we have to handle this aspect carefully}'' (P12). Also, P14 highlighted concerns about technological devices like home security cameras, using them as an example to illustrate students' reluctance to share data. He stated, ``\textit{Some students might refuse to consent to data collection due to concerns about information leakage. Especially as they grow older and become more aware of privacy issues, it is likely that some will decide against participating.}''

Further, the possibility of student data leakage was also raised as a significant concern: ``\textit{It could be risky if student records get leaked}'' (P17). Further, some participants mentioned the potential negative impact on students' comfort and autonomy. They said that students might feel constantly monitored, with the perception that everything is being recorded, assessed, and potentially censored. In this regard, P12 mentioned ``\textit{Everything about students becomes data. So, there might be situations where someone doesn’t want to give their data but is required to.}''

\section{Discussion}
Our study uncovered potential applications and key considerations for integrating AI into K-12 PE settings. In this section, we reflect on the findings and discuss considerations for designing and implementing AI systems to address the unique demands of PE.

\subsection{Multidimensional Roles of AI in PE}
Our findings revealed that PE teachers recognize the potential of AI to significantly enhance their classes. Participants envisioned AI attaining the multi-dimensional educational goals of K-12 PE, aligning with previous research on AIED ~\cite{lee2021applying, miguel2017foreword, lytle2018physical, santos2017assessment, roure2018exploring, guliherme2017ai, park2017students, xia2022systematic}. However, teachers consistently emphasized that AI should act as a supportive tool to complement rather than replace human educators. This reflects a consensus that AI alone cannot fully address the complex and multifaceted challenges of PE, a perspective echoed in existing literature ~\cite{ghamrawi2023exploring}.

Specifically, analysis of participants' ideas, sketches, and storyboards uncovered four primary supporting roles for AI in PE: operational assistant, personal trainer, group coach, and evaluator. These roles reflect the potential of AI to assist with administrative tasks, personalized instruction, balanced team activities, and performance assessments. This comprehensive vision spanned from administrative support to personalized instruction and assessment, demonstrating a broader range of AI applications in PE than typically found in existing AIED research that focused primarily on student-oriented solutions ~\cite{crompton2021potential, crompton2020psychological, chen2020artificial, vanlehn2011relative, ma2014intelligent}.  
These insights from PE teachers reveal a nuanced understanding of classroom dynamics and educational needs that may not be apparent to AI developers or researchers lacking direct PE teaching experience.

In particular, participants discussed the concept of an \textbf{Operational Assistant}, emphasizing its potential to optimize instructional time by streamlining administrative and operational tasks. This suggests that AI's utility in PE extends beyond direct student engagement, offering significant value for enhancing the learning environment and ensuring safety during physical activities. These findings suggest that AI applications in classroom settings should not be confined to knowledge instruction but should also encompass a broader range of the educational ecosystem.


Further, participants highlighted limited instructional time as a key challenge and proposed a \textbf{Personal Trainer} role for AI to provide personalized feedback. Personalized instruction has been a focal point in K-12 education, primarily providing problems and interpretations aligned with individual learning levels ~\cite{drumheller1971handbook, slavin1984team, subban2006differentiated, tomlinson1999mapping}. Personalized feedback has demonstrated improvements in knowledge comprehension and acquisition, leading to enhanced attitudes ~\cite{hwang2012development}, engagement ~\cite{arroyo2014multimedia}, and academic performance ~\cite{zhang2020understanding, walkington2013using} in other subjects. However, PE, as a non-sequential subject that covers diverse content without a specific order or hierarchy, presents unique challenges. In particular, students in PE exhibit widely varying levels of skills and engagement based on their prior experiences, physical conditions, and abilities ~\cite{bertills2019inclusive}. Considering this diversity, the types of feedback required in PE may extend beyond skill and knowledge acquisition to include posture correction, performance improvement, and mental support. 


Furthermore, participants envisioned AI serving as a \textbf{Group Coach} in team-based physical activities, with a particularly pronounced need for balanced team formation.
PE's incorporation of sports results in a distinctive educational environment that is simultaneously competitive and cooperative. This dual nature makes balanced grouping a crucial factor in students' social skill development ~\cite{metzler2017instructional}. While group formation is important across various subjects for discussions or team projects, considering students' experiences, interests, and diversity ~\cite{santangelo2012teacher, chapman1995designing}, solutions to address the unique challenges PE faces in this regard have been limited. AI's potential role as a Group Coach might offer solutions to this challenge, potentially considering multiple factors such as physical abilities, social dynamics, and individual preferences to form optimally balanced teams.

Lastly, participants envisioned AI taking on the role of an \textbf{Evaluator}, collecting and assessing students' progress from multiple angles. This approach aimed to evaluate students' sports knowledge, athletic performance, and attitudes, considering both quantitative and qualitative aspects to track their overall growth. While such thorough assessment is critical, it significantly increases teacher workload, highlighting the need for AI-based solutions. Existing AI assessment systems, such as those used for essay grading or problem-solving assignments, primarily focus on knowledge evaluation for formative and summative purposes ~\cite{zhu2020effect}. However, PE requires assessments that integrate skills, attitudes, and knowledge, presenting a unique and valuable opportunity for AI application. 

\subsection{Considerations for AI Implementation in PE}
Our group ideation workshops with K-12 PE teachers highlighted promising opportunities for integrating AI into PE classes. However, several critical considerations emerged that need to be considered to ensure the successful and safe implementation of AI technologies in real-world PE settings.

\subsubsection{Ensuring Student Data Security and Privacy}
A key concern centers around the legal and ethical implications of student data collection. Since the introduction of digital devices in education, student data privacy has been a longstanding issue ~\cite{kwapisz2024privacy}. Participants voiced apprehension about potential data leaks and the need for clear student consent, particularly given the sensitive nature of the biometric and physical activity data collected in PE.

This concern aligns with existing research, which shows that students often share personal data without fully understanding the associated risks ~\cite{solove2012introduction}, even as they consistently express concerns about privacy ~\cite{kwapisz2024privacy}. Moreover, students value transparency regarding why their data is being collected and how it will be used ~\cite{ngoon2023instructor, slade2014student, sun2019s}. Without addressing these privacy concerns, the implementation of AI in PE could create an environment of constant monitoring, potentially leading to feelings of restriction or self-censorship. This may inhibit the free interaction and collaboration that PE classes aim to foster.

To mitigate this, it would be essential to provide students with full disclosure regarding the data being collected, including the methods, purposes, and scope of access as suggested by prior work ~\cite{royal2019protecting}. This transparency would empower students to actively participate in decisions regarding their data ~\cite{prinsloo2022answer}, ensuring that AI enhances their learning experience rather than undermining the open, collaborative nature of PE.

\subsubsection{Reducing Risks of Over-reliance on AI in PE}
Participants expressed significant concern about the potential for students to over-rely on AI instructions in PE classes. They worried that excessive dependence on AI could reduce student autonomy and diminish the teacher's role. Specifically, participants argued that if AI assumes too much responsibility for providing instructions and feedback, it might limit the teacher's direct involvement and guidance, weakening the teacher's influence in the educational process. While participants acknowledged the potential for automation in certain contexts—such as AI serving as an Evaluator to assist with assessment tasks—they emphasized the need for teachers to retain control during class time, with AI playing a supportive role.

This hesitation reflects a lack of consensus on the appropriate balance between AI and traditional student-teacher interactions in PE. Without clear guidelines, both students and teachers risk over-relying on AI, treating its judgments as definitive  ~\cite{diberardino2023anti, araujo2020in}. Such over-reliance could compromise the nuanced understanding and professional judgment that experienced educators bring to student assessment and development.

To mitigate these risks, it would be crucial to involve both students and teachers in discussions about the appropriate level of AI integration in PE classes. This collaborative approach would help ensure that AI technologies are implemented in ways that maintain educational transparency, accountability, and the integrity of student-teacher relationships.

\subsubsection{Narrowing Technological Proficiency and Motivation Gaps among PE Teachers}
Participants expressed significant concerns about disparities in teachers’ technological proficiency and motivation to integrate AI into their classes. These gaps could result in inconsistent instructional quality, with AI-proficient teachers delivering more effective lessons while those less adept with technology might struggle to utilize AI tools effectively. This concern was particularly pronounced among teachers who had faced difficulties during the rapid digital transition necessitated by COVID-19. Their experiences with technical challenges during the shift to online teaching heightened apprehensions about encountering similar obstacles in the adoption of AI. 

These findings suggest a clear need for a gradual introduction of AI technologies in PE. This approach should be supported by well-designed PE teacher training programs that focus on building both technical proficiency and confidence in AI utilization. Such initiatives would help ensure that all teachers, regardless of their initial skill and motivation levels, can effectively incorporate AI into their teaching practices, promoting equitable and consistent educational outcomes across PE classes.

\subsubsection{Securing Institutional Support and Addressing Resource Constraints}
The practical implementation of AI in PE classes extends beyond individual teacher capabilities, as our findings revealed that successful AI adoption would be heavily dependent on the willingness and economic support of educational institutions ~\cite{landi2021physical, cothran2014classroom}. However, the relatively low-priority status of PE in many school curricula might make this process challenging. 

Securing institutional support is essential for AI integration, which requires clearly demonstrating the benefits and necessity of AI in PE settings to justify the investment. Given the financial constraints many schools face, it would be useful to present a compelling case for allocating resources to AI in PE. For example, small-scale AI implementation attempts using already available data and AI technology could be initiated in the early stages. For instance, a previous study demonstrated the potential of using basic parameters such as gender, height, weight, age, and heart rate to tailor exercise plans for individuals with lower fitness levels ~\cite{chen2021ai, Wilder2006physical}. These kinds of pilot projects could serve as a proof-of-concept to persuade institutions of AI's value in PE without necessitating extensive new data collection or infrastructure. Moreover, these small-scale projects could help identify potential challenges and refine implementation strategies before full-scale adoption, ensuring more efficient use of resources. This gradual, evidence-based approach to AI integration in PE would be key to securing the institutional support necessary for widespread implementation, even in the face of competing priorities within educational systems.

\subsection{Limitation and Future Work}
While our study highlighted AI's potential in K-12 PE, several limitations should be considered, along with directions for future research. 

Firstly, the study design inherently guided participants to consider AI as the solution to challenges in their PE classes. By structuring the focus group ideation activities around AI integration, we might have constrained participants from exploring more fundamental or non-technological solutions, such as teacher training, curriculum adjustments, or policy changes. This enforced focus on AI might have led participants to overlook simpler or potentially more effective strategies that do not involve technology. Therefore, future research will need to adopt a more open-ended approach to ideation, allowing participants to explore both technological and non-technological solutions for a more balanced perspective.

Second, although the participants represented a wide range of teaching experience from 1 to 25 years, they all identified as South Korean. This lack of diversity in terms of nationality and cultural background could impact the generalizability of the findings. Additionally, the study predominantly focused on teachers from urban areas in South Korea, which means that the perspectives and experiences of teachers from rural areas might not be adequately represented. Another limitation of this study lies in the fact that none of the participating teachers had experience teaching multicultural students. Future studies will need to recruit more diverse PE teachers from diverse backgrounds, including those from different regions, cultures, and educational contexts. 

Moreover, the participants in our study displayed a notably proactive and open-minded attitude toward the topic of AI. Our participants' proactive approach towards AI integration in PE might not reflect the apprehension or skepticism towards AI among other educators \cite{christian2024impact}. Additionally, because of the structure of the focus group ideation activities, participants might have focused more on the positive aspects of AI technology in PE classrooms than on the potential problems. Future research could incorporate more critical and balanced discussions, perhaps by including scenarios or case studies that highlight potential drawbacks of AI integration.

Lastly, while our study primarily focused on teachers' perspectives, the introduction of AI into PE settings is indeed a complex process that involves multiple stakeholders, each with their own priorities and concerns. Thus, future research should aim to involve a wider range of stakeholders at each stage of the AI adoption process. This could include students of various ages and abilities, parents and guardians, school administrators, PE educators, and AI developers.

\section{Conclusion}
This study explores AI's potential in the context of K-12 PE, identifying four roles for AI—operational assistant, personal trainer, group coach, and evaluator—that address instructional and operational challenges unique to PE. By engaging secondary PE teachers, our findings highlight how AI in the PE context can not only enhance instruction but also improve a wide range of activities, such as class management, personalized feedback, facilitating balanced team activities, and streamlining performance assessments. The findings also emphasize the importance of addressing practical and ethical considerations, such as data privacy, feedback reliability, and workload concerns, to ensure AI systems meet educators' needs. Shifting the focus from STEM subjects to PE, this work highlights the need for equitable access to AI innovations across educational domains and underscores the pivotal role of teachers in shaping AI applications to meet pedagogical goals. We hope these insights provide a foundation for future research aimed at developing ethical, scalable, and context-sensitive AI solutions for K-12 education.

\begin{acks}
We are deeply grateful for our participants and reviewers who significantly contributed to this work. We would also like to thank Dr. Jooeun Ahn for his feedback on the earlier version of this paper. This work was supported by the New Faculty Startup Fund from Seoul National University (\#200-20230022) and the Undergraduate Research Learner (URL) program of Information Science and Culture Studies at Seoul National University. 
 
\end{acks}

\bibliographystyle{ACM-Reference-Format}
\bibliography{ref}
\end{document}